\documentclass[%
 reprint,
superscriptaddress,
 amsmath,amssymb,
 aps,
]{revtex4-2}

\usepackage{graphicx}% Include figure files
\usepackage{dcolumn}% Align table columns on decimal point
\usepackage{bm}% bold math
\usepackage[scheme=plain]{ctex}
\usepackage{tikz-cd}
\usepackage{tikz}
\usetikzlibrary{calc}
\usepackage{booktabs}
\usepackage{siunitx}
\usepackage{multirow}
\newcommand{\gluonFracJetA}{$10.3\%$}
\newcommand{\gluonFracJetB}{$26.3\%$}
\newcommand{\gluonFracJetC}{$63.2\%$}

\newcommand{\gluonFracJetAAlt}{$10.9\%$}
\newcommand{\gluonFracJetBAlt}{$24.7\%$}
\newcommand{\gluonFracJetCAlt}{$56.1\%$}

\newcommand{\aveXiJetA}{$2.51^{+0.11}_{−0.10} (\text{stat}) \pm 0.19 (\text{syst})$}
\newcommand{\aveXiJetB}{$2.33^{+0.10}_{−0.10} (\text{stat}) \pm 0.16 (\text{syst})$}
\newcommand{\aveXiJetC}{$2.84^{+0.05}_{−0.05} (\text{stat}) \pm 0.082 (\text{syst})$}

\newcommand{\RgqCentral}{1.21}
\newcommand{\RgqStat}{0.18}
\newcommand{\RgqSys}{0.26}

\begin{document}

% \preprint{APS/123-QED}

\title{Measurement of $\Xi^-/\bar{\Xi}^{+}$ production in jets from $Z$ boson decays using DELPHI open data}%
\thanks{Project supported by the National Natural Science Foundation of China (Grant No.~12342502).}%

\author{Zhouyue Fan (范周越)}%
\affiliation{Institute of High Energy Physics, Chinese Academy of Sciences, Beijing, 100049, Beijing, China}%
\affiliation{University of Chinese Academy of Sciences, Beijing, 100049, Beijing, China}%

\author{Yuzhi Che (车逾之)}%
\email{cheyuzhi@ihep.ac.cn}%
\thanks{Corresponding author.}%
\affiliation{Institute of High Energy Physics, Chinese Academy of Sciences, Beijing, 100049, Beijing, China}%
\affiliation{China Center of Advanced Science and Technology}%

\author{Xinchou Lou (娄辛丑)}%
\affiliation{Institute of High Energy Physics, Chinese Academy of Sciences, Beijing, 100049, Beijing, China}
\affiliation{University of Chinese Academy of Sciences, Beijing, 100049, Beijing, China}%
\affiliation{Center for High Energy Physics, Henan Academy of Sciences, Zhengzhou, 450046, Henan, China}%
\affiliation{University of Texas at Dallas, Richardson, 75083, Texas, USA}%

\author{Zhaoru Zhang (张照茹)}%
\affiliation{Institute of High Energy Physics, Chinese Academy of Sciences, Beijing, 100049, Beijing, China}%

\author{Jinfei Wu (吴金飞)}%
\affiliation{Institute of High Energy Physics, Chinese Academy of Sciences, Beijing, 100049, Beijing, China}%

\author{Gengyuan Zhang (张耕源)}%
\affiliation{Institute of High Energy Physics, Chinese Academy of Sciences, Beijing, 100049, Beijing, China}%
\affiliation{University of Chinese Academy of Sciences, Beijing, 100049, Beijing, China}%

\author{Gang Li (李刚)}%
\affiliation{Institute of High Energy Physics, Chinese Academy of Sciences, Beijing, 100049, Beijing, China}%
\affiliation{University of Chinese Academy of Sciences, Beijing, 100049, Beijing, China}%

\author{Yanping Huang (黄燕萍)}%
\affiliation{Institute of High Energy Physics, Chinese Academy of Sciences, Beijing, 100049, Beijing, China}%
\affiliation{University of Chinese Academy of Sciences, Beijing, 100049, Beijing, China}%

\author{Manqi Ruan (阮曼奇)}%
\affiliation{Institute of High Energy Physics, Chinese Academy of Sciences, Beijing, 100049, Beijing, China}%
\affiliation{University of Chinese Academy of Sciences, Beijing, 100049, Beijing, China}%

\date{\today}% It is always \today, today,

\begin{abstract}
 The production rates of $\Xi^{-}/\bar{\Xi}^{+}$ baryons in energy-ranked jets produced in $Z\to\text{hadrons}$ decays are measured using $3.2$ million hadronic $Z$ events recorded by the DELPHI experiment. Jets are reconstructed using the Durham algorithm with $y_{\text{cut}}=0.005$. Quark- and gluon-enriched jet samples are obtained by ranking the jet energies in three-jet events. The softest jet are found to produce fewer $\Xi^{-}/\bar{\Xi}^{+}$ and less energetic baryons than the other jets. The ratio of $\Xi^{-}/\bar{\Xi}^{+}$ production rates in gluon and quark jets, each normalized to the corresponding mean charged-particle multiplicity, is measured to be $\RgqCentral \pm \RgqStat~\mathrm{(stat.)} \pm \RgqSys~\mathrm{(syst.)}$. The result is consistent with the JETSET expectation and the OPAL measurements of $K_S^0$ and $\Lambda$ productions in $Z$ decays. This study presents the first measurement of the gluon-to-quark production ratio for baryons containing two $s$ quarks, providing new insights into strange-quark production and hadronization. Future $e^{+}e^{-}$ colliders such as CEPC and FCCee will provide much larger $Z$-boson samples and will allow far more precise studies of the subject.
\end{abstract}

\maketitle

%\tableofcontents

\section{Introduction}

% Quantum chromodynamics (QCD) describes the interactions between quarks and gluons and provides the theoretical framework for understanding hadron formation. Over the past three decades, perturbative QCD has
% been well established and has successfully described processes at high
% energy scales. However, the theoretical description of non-perturbative
% effects still relies on phenomenological assumptions and artificial
% parameterizations. Hadronization is an example of such a
% non-perturbative process and is usually modeled within semi-classical
% frameworks, represented by the Lund string~\cite{Andersson:1997xwk} and
% cluster~\cite{Webber:1983if} fragmentation models. Although these
% models successfully describe many inclusive distributions of hadron production,
% their predictive power is more limited for the production of identified hadrons and their correlations. In particular, strangeness production and baryon-number compensation remain strongly dependent on phenomenological parameters tuned to data.

In the prediction of the Quantum Chromodynamics (QCD) theory, 
gluon jets are expected to exhibit stronger color radiation
in the perturbative parton cascade and to contain larger particle
multiplicities~\cite{Capella:1999ms, Bethke:2004bp} in hadronic $Z$ decays. This general
enhancement is largely reflected in inclusive charged-particle
production.
For identified hadron species, the observed differences
between quark and gluon jets also encode properties of
non-perturbative QCD. Earlier studies investigated $K_S^0$ and $\Lambda$ production in energy-ranked jets and derived the ratios of their production rates in gluon and quark jets relative to the mean charged-particle multiplicity,
\begin{equation}
\mathcal{R}^{h}_{g/q} \equiv \frac{\langle N_{h}\rangle_{g}}{\langle N_{\text{ch}}\rangle_{g}}\left/\frac{\langle N_{h}\rangle_{q}}{\langle N_{\text{ch}}\rangle_{q}}\right.,
\label{eq:Rgq}
\end{equation}\\
where $\langle N_{h}\rangle_{g}$ and $\langle N_{h}\rangle_{q}$ are the average yields of hadron type $h$ in gluon and quark jets, respectively, while $\langle N_{\text{ch}}\rangle_{g}$ and $\langle N_{\text{ch}}\rangle_{q}$ denote the mean charged-particle multiplicities in gluon and quark jets.
The results showed that identified
strange-hadron production in gluon jets cannot be fully understood
through a general scaling of inclusive charged-particle multiplicity in gluon jets~\cite{OPAL:1998izp}. 
Furthermore, measurements of single-strange baryon--antibaryon
correlations indicate that baryon-number and strangeness compensation
in QCD depend on the baryon species and rapidity
separation~\cite{OPAL:1998tlp,OPAL:2009hyp,DELPHI:1997fmy}. In this
context, the production of doubly-strange baryons, $\Xi^{-}$, in jets
intrinsically involves both the strange hadron production and local compensation of baryon 
number, making it a sensitive probe of non-perturbative QCD.
The production of $\Xi^{-}$ baryons has been experimentally investigated in inclusive $Z$ decays by OPAL~\cite{OPAL:1992asw, OPAL:1996gsw} and DELPHI~\cite{DELPHI:1995dso, DELPHI:2006pom}, but its dependence on the jet environment remains unexplored. Three-jet hadronic $Z$ events provide experimentally accessible quark- and gluon-enriched jet samples, enabling a differential study of $\Xi^{-}$ production in distinct jet categories.

% investigated at different energy scales. Near the threshold, the $e^{+}e^{-}\to
%     \Xi^{-}\bar{\Xi}^{+}$ cross-section has been measured by
% BESIII~\cite{BESIII:2019cuv,BESIII:2020ktn}. In this energy region, the $\Xi^{-}$ production is
% described in the context of Born cross-section, electromagnetic form factors, and effects of intermediate resonance.
% As the energy scale increasing, the $\Xi^{-}$ production is discussed using statistic language.
% The inclusive $\Xi^{-}$ production has been measured near the $\Upsilon(4S)$ resonance 
% by Belle~\cite{Belle:2017caf},
% and in hadronic $Z$ decays by OPAL~\cite{OPAL:1992asw, OPAL:1996gsw} and
% DELPHI~\cite{DELPHI:1995dso, DELPHI:2006pom}. ALICE studied
% multi-strange baryon production, including $\Xi^{-}$ and $\Omega$
% baryons, in $pp$ collisions at the center-of-mass energy of $7$ and
% $13\mathrm{\,TeV}$~\cite{ALICE:2012yqk, DerradideSouza:2016kfn}.
% These two views provide complementary descriptions for the $\Xi^{-}$ production mechanism.
% In the latter context, in addition, hadronic
% $Z$ decays offer a well-defined initial state, together with the
% absence of the underlying events. The three-jet hadronic $Z$ events
% provide experimentally accessible quark- and gluon-enriched jet
% samples, enabling a more differential test of $\Xi^{-}$ production,
% which remains unexplored.

In 2024, the DELPHI data set and the full simulation and reconstruction
software stack were released to the global high-energy-physics
community~\cite{DELPHI:OpenData2024}, making it possible to investigate further into the 
strange-baryon production in $Z$ decays.

This paper reports a measurement of  $\Xi^{-}/\bar{\Xi}^{+}$ production in two- and
three-jet events using DELPHI data collected during 1992--1995. The
production rates are reported as functions of jet energy and jet energy
rank. The $\xi=-\ln(p_{\Xi}/E_{\text{beam}})$ spectra are compared
among energy-ranked jet samples. 
For three-jet events, the
generator-modeled gluon fractions of the energy-ranked jets are used to
infer the ratio of $\Xi^{-}/\bar{\Xi}^+$ production rates per mean charged-particle multiplicity in gluon and quark jets, $\mathcal{R}^{\Xi^{-}}_{g/q}$.

The rest of this paper is organized as follows.
Section~\ref{sec:xi_rec} summarizes the DELPHI detector, the open data
samples, and the corresponding simulated samples used in the analysis.
Section~\ref{sec:jet_xi_reco} describes the jet reconstruction, the
energy-ranked jet classification, the generator-level gluon-fraction
estimate, and the reconstruction and selection of
$\Xi^{-}/\bar{\Xi}^{+}$ candidates. The $\Xi^{-}/\bar{\Xi}^{+}$ yield
extraction, efficiency correction, uncertainty evaluation, and the
resulting measurements in energy-ranked jets are presented in
Section~\ref{sec:analysis}. The main findings are summarized in the final
section.
Unless stated otherwise, $\Xi^{-}$ in the following denotes the sum of
$\Xi^{-}$ and its charge-conjugate state $\bar{\Xi}^{+}$.

\section{The DELPHI detector, data and simulation samples}\label{sec:xi_rec}

The DELPHI detector and its performance are described in detail in
Refs.~\cite{DELPHI:1990cdc, DELPHI:1995dsm}. From the beam pipe
outwards, the central detector consisted of the Vertex Detector (VD),
the Inner Detector (ID), the Time Projection Chamber (TPC), the Outer
Detector (OD), the Ring Imaging Cherenkov counter (RICH), the
High-density Projection Chamber electromagnetic calorimeter (HPC), the
Hadron Calorimeter (HAC) instrumented in the iron return yoke, and the
barrel muon chambers. The whole tracking system operated inside a
magnetic field of $1.2\,\mathrm{T}$, parallel to the beam axis. During
the 1991--1993 running period, the VD consisted of three layers of
single-sided silicon microstrip detectors at radii of approximately
$6.3$, $9$, and $11\,\mathrm{cm}$; from 1994 onward, the use of
double-sided silicon detectors allowed three-dimensional
impact-parameter reconstruction, improving secondary-vertex
reconstruction and heavy-flavour tagging~\cite{DELPHI:1995dsm}.

This study uses the $Z$-decay data collected in 1992--1995. The
sample nicknames and Digital Object Identifiers (DOIs) are listed in
Table~\ref{tab:nickname}. The corresponding MC samples of inclusive
$e^{+}e^{-}\to Z\to q\bar{q}$ events, produced with the DELPHI
built-in generator~\cite{DELPHI:MC} and released with the data set, are
used to interpret the data and estimate the selection efficiency for
$\Xi^{-}$ candidates. These samples are also listed in
Table~\ref{tab:nickname}.
These samples are stored in the original DELPHI DST data format.
A dedicated tool, adapted from the software described in Ref.~\cite{Zhang:2025nlf}, is used to extract the information relevant to the analysis from the DST files and store it in files in ROOT format~\cite{Brun:1997pa}.

\begin{table}[htbp!]
    \centering
    \caption{Nicknames and Digital Object Identifiers (DOIs) of the data sets and official simulation samples used in this analysis.}
    \label{tab:nickname}
    \begin{tabular}{llll}
        \toprule
        Type & Year & Sample                                & DOI                                            \\
        \midrule
        Data & 1992 & short92\_e2                           & \cite{delphi_short92_e2}                       \\
        Data & 1993 & short93\_d2                           & \cite{delphi_short93_d2}                       \\
        Data & 1994 & short94\_c2                           & \cite{delphi_short94_c2}                       \\
        Data & 1995 & short95\_d2                           & \cite{delphi_short95_d2}                       \\ \midrule
        MC   & 1992 & sh\_kk2f4146qqpy\_e91.25\_c92\_2l\_e2 & \cite{delphi_sh_kk2f4146qqpy_e91_25_c92_2l_e2} \\
        MC   & 1993 & sh\_kk2f4146qqpy\_e91.25\_c93\_2l\_d2 & \cite{delphi_sh_kk2f4146qqpy_e91_25_c93_2l_d2} \\
        MC   & 1994 & sh\_kk2f4146qqpy\_e91.25\_l94\_2l\_c2 & \cite{delphi_sh_kk2f4146qqpy_e91_25_l94_2l_c2} \\
        MC   & 1995 & sh\_kk2f4146qqpy\_e91.25\_c95\_1l\_d2 & \cite{delphi_sh_kk2f4146qqpy_e91_25_c95_1l_d2} \\
        \bottomrule
    \end{tabular}
\end{table}

All events are required to pass the official "Team 4" selection encoded in
the open data as a tag, after which approximately $3.23\times
    10^6$ events remain. In each event, reconstructed particles are selected according to the official criteria~\cite{DELPHI:1995dsm, DELPHI:2024:skelana-manual}.

\section{Jet clustering and $\Xi^{-}$ reconstruction}\label{sec:jet_xi_reco}
To investigate $\Xi^{-}$ production in jets, the
$e^{+}e^{-}$ $k_{\perp}$ clustering algorithm, known as the
Durham algorithm~\cite{Catani:1991hj}, is implemented using the
\textsc{Fastjet}~\cite{Cacciari:2011ma} package.
All charged and neutral reconstructed particles satisfying the requirements described in
Section~\ref{sec:xi_rec} are used as jet-clustering inputs.
In the Durham algorithm, the
distance between input objects $i$ and $j$ is defined as
\begin{equation}
    y_{ij} = \frac{2 \operatorname{min}(E_i^2, E_j^2) (1-\cos\theta_{ij})}{Q^2},
    \label{eq:y}
\end{equation}
where $E_{i}$ and $E_{j}$ are the object energies, $Q$ is the
center-of-mass energy, and $\theta_{ij}$ is the opening angle between the two objects.
Following Ref.~\cite{OPAL:1998izp}, the jet resolution parameter is set to $y_{\text{cut}} = 0.005$.

The jet energy is defined as the sum of the energies of the
reconstructed particles assigned to the jet, with charged particles
treated as pions and neutral particles treated as photons. In
two- and three-jet events, the reconstructed jets are ordered by energy and
denoted as the most energetic jet (\textit{\textbf{Jet 1}}), the second-most
energetic jet (\textit{\textbf{Jet 2}}), and the third-most energetic jet (\textbf{\textit{Jet 3}}).

The gluon fraction of each energy-ranked jet category is estimated
from the MC samples using generator-level parton information. A list of
generator-level reference partons is constructed from the
primary partons whose parent is the generated $Z$ boson. 
The most energetic generated gluon in the event is included.
Reconstructed jets are matched to generator-level partons based on the smallest angular separation between their momentum directions.
The gluon fraction is defined as the fraction
of jets in the category that are matched to a gluon.
As a result, the average gluon fractions of \textbf{\textit{Jet 1}}, \textbf{\textit{Jet 2}}, and \textbf{\textit{Jet 3}} are \gluonFracJetA, \gluonFracJetB, and \gluonFracJetC, respectively.

\begin{figure}[htb!]
    \centering
    \includegraphics[width=0.7\linewidth]{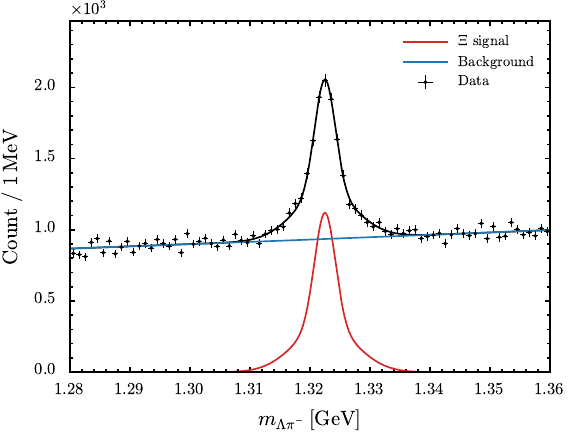}\quad
    \caption{
        The invariant mass distribution of the selected $\Xi^{-}$ candidates from all the data samples listed in Table~\ref{tab:nickname}.
        The spectrum is fitted with a double-Gaussian function (red curve) and a third-order polynomial function (blue curve).
    }
    \label{fig:m_xi}
\end{figure}

The $\Xi^{-}$ candidates are reconstructed via the cascade decay $\Xi^{-}\to \Lambda \pi^{-}$,
followed by $\Lambda\to p\pi^{-}$.
$\Lambda$ candidates are reconstructed from oppositely charged particle pairs ($V^{0}$ candidates)
originating from neutral particles decaying in flight, using the standard DELPHI $V^{0}$ algorithm~\cite{DELPHI:1995dsm}.
All $V^{0}$ candidates are treated as $\Lambda$ candidates and are combined with additional charged tracks to form $\Xi^{-}$ cascade decays.
An external
package for vertex fitting~\cite{Xu:2009zzg} is used to reconstruct the two vertices in the $\Xi^{-}$
cascade. 
The higher-momentum daughter track of the $\Lambda$ candidate is assigned the proton mass, while the other daughter is assigned the pion mass.
The momenta of the three charged daughter tracks are corrected using the corresponding $\Xi^{-}$ and $\Lambda$ decay vertices as their points of origin.

For $\Xi^{-}$ candidate selection, the two pion tracks are
required to have the same charge. The $\Lambda$ candidates must satisfy $|m_{\Lambda}-1116.3~\mathrm{MeV}|<4.5~\mathrm{MeV}$. The
$\chi^{2}$ probabilities for both constraints in the
$\Xi^{-}\to\Lambda\pi^{-}$ vertex fit are required to be greater than
$0.1\%$. To further suppress combinatorial background, the signed decay radius of each $\Xi^{-}$ candidate in the $R\phi$ plane is required to satisfy
\begin{equation}
    R_\Xi =
    \left.
    \begin{cases}
        +\,r_\Xi & \text{if } r_\Lambda > r_\Xi,    \\
        -\,r_\Xi & \text{if } r_\Lambda \leq r_\Xi,
    \end{cases}
    \right\}
    > 0.4\,\mathrm{cm},
\end{equation}
where $r_\Xi$ and
$r_\Lambda$ are the transverse distances of the $\Xi^{-}\!\to\!\Lambda\pi^{-}$ and
$\Lambda\!\to\!p\pi^{-}$ decay vertices from the beam spot, respectively.
Because the DELPHI detector has limited acceptance for $\Xi^{-}$ baryons outside the range $1.0< \xi \equiv -\ln(p_{\Xi}/E_{\text{beam}}) < 4.0$,
only candidates within this range are used in this analysis.
The resulting invariant-mass spectrum of the $\Xi^{-}$ candidates in all $Z$-decay events is shown in Fig.~\ref{fig:m_xi}.
\footnote{Because the magnetic-field conditions are not fully known, the $\Xi^{-}$ candidate mass distributions are biased.
Following the calibration used in the official study~\cite{DELPHI:2006pom}, a residual mass-scale correction
is applied to the $\Xi^{-}$ mass in data. The correction is calibrated using the mass shifts of $\Lambda$ and $K_{S}^{0}$ candidates, and the distribution in the MC samples is artificially shifted to match the nominal $\Xi^{-}$ mass.}

Each reconstructed $\Xi^{-}/\bar{\Xi}^{+}$ candidate is assigned to
a jet according to its distance from the reconstructed jets.
The quantity $y_{\Xi\,\text{jet}}$, defined in Eq.~\ref{eq:y}, is calculated using
the four-momenta of each $\Xi^{-}$ candidate and jet. If the minimum
value of $y_{\Xi\,\text{jet}}$ for a given $\Xi^{-}$ candidate is smaller than
$y_{\text{cut}}$, the candidate is assigned to the corresponding jet.

\section{Average $\Xi^{-}$ production in energy-ranked jets}\label{sec:analysis}

This analysis begins with the selection of hadronic events.
All jets in each event must have polar angles in the range $30^\circ<\theta_{\mathrm{jet}}<150^\circ$.

Jets in 2- and 3-jet events are grouped into bins of reconstructed jet energy. 
The average $\Xi^{-}$ production per jet is calculated as
\begin{equation}
    \left\langle N_{\Xi}\right\rangle
    =
    \frac{1}{N_{\text{jet}, k}}
    \frac{N^{\Xi^{-}}_{\text{obs},k}}
         {\epsilon_{k}
          f_{\text{acc},k}},
    \label{eq:observable}
\end{equation}
where $N_{\text{jet}, k}$ is the number of reconstructed jets included in the $\Xi^{-}$ production measurement.
The number of observed $\Xi^{-}$ candidates ($N_{\text{obs},k}$) is obtained from fits to the $\Xi^{-}$ mass spectra in the corresponding jets.
The quantities $\epsilon_{k}$ and $f_{\text{acc},k}$ are the selection efficiency and acceptance factor, respectively, estimated from the MC samples.

When fitting the $\Xi^-$ mass spectrum,
the signal probability density function (PDF) is modeled using
kernel density estimation (KDE) with a Gaussian kernel based on the MC simulation.
The KDE bandwidth follows the adaptive prescription of Ref.~\cite{Abramson:1982kde}. 
The combinatorial background is modeled using wrong-sign control candidates.
These candidates are formed from $\Lambda \pi^{+}$ and $\bar{\Lambda}\pi^{-}$ 
combinations and are passed through the same reconstruction procedure and selection requirements as the right-sign $\Xi^{-}$ candidates.
The wrong-sign candidates are then used to construct the background KDE template.
The obtained signal yields are used as the raw inputs to the data correction described below.

% \subsection{Efficiency correction}

The $\epsilon_{k}$ and $f_{\text{acc},k}$ in Eq.~\ref{eq:observable} are estimated using MC simulations.
In the MC samples, generated $\Xi^{-}$ baryons are assigned to jets using the same Durham-distance criterion as that used in the jet-clustering algorithm. The generated $\Xi^{-}$ baryons are then matched to the reconstructed $\Xi^{-}$ candidates using MC information.
For each energy-ranked jet, the efficiency $\epsilon_{k}$ is defined as the ratio of selected
$\Xi^{-}$ candidates matched to truth $\Xi^{-}$ baryons to the total number of generated truth $\Xi^{-}$ baryons.
The acceptance factor $f_{\text{acc}, k}$ is defined as the ratio of generated $\Xi^{-}$ baryons within $1.0<\xi<4.0$ to all generated $\Xi^{-}$ baryons assigned to the corresponding jet category.

% \subsubsection{Uncertainties}\label{sec:unc}
\paragraph*{\textbf{Uncertainties}} The statistical uncertainty includes contributions from the fitted signal yield in data
and the finite MC statistics entering the efficiency and acceptance correction.
These components are propagated through the correction procedure.

The systematic uncertainty from the $\Xi^{-}$ selection is evaluated by repeating the
measurement with varied selection requirements. The
$m_{\Lambda}$ window is varied from the nominal $4.5\mathrm{\,MeV}$ to
$4.0\mathrm{\,MeV}$ and $5.0\mathrm{\,MeV}$. The requirement on the signed $\Xi^{-}$
decay radius in the $R\phi$ plane is varied from the nominal $0.4\mathrm{\,cm}$ to
$0$, $0.5\mathrm{\,cm}$, and $2.5\mathrm{\,cm}$. For each source,
the largest variation with respect to the nominal result is
taken as the corresponding systematic uncertainty.

The systematic uncertainty associated with the signal yield extraction model is estimated with an alternative method.
In this procedure, the background in each $\Xi^{-}$ mass spectrum is modeled with wrong-sign control samples, and its normalization is determined by fitting the spectrum outside a
$\pm 10\mathrm{\,MeV}$ window around the $\Xi^{-}$ mass. 
The signal yield is then obtained by subtracting the normalized background contribution from the data within the $\pm 10\mathrm{\,MeV}$ signal window.
Half of the difference between this yield and the nominal fit result is assigned as the fit-model systematic uncertainty.

The acceptance factor used to extrapolate from $1.0<\xi<4.0$ to the full kinematic range is derived from the JETSET generator and is therefore model dependent.
Following the previous study~\cite{DELPHI:2006pom}, a relative uncertainty of $\pm 50\%$ is assigned to the amount of compensation.
The total systematic uncertainty is obtained by adding the individual systematic components in quadrature.

\begin{table*}[htbp!]
  \centering
  \caption{Average \ensuremath{\Xi} production per jet in two- and three-jet events. The central observable is computed as $\langle N_{\Xi}\rangle=N_{\text{obs}}/(\epsilon f_{\text{acc}}N_{\text{jet}})$ from the fitted signal yield ($N_{\text{obs}}$), selection efficiency ($\epsilon$), acceptance factor ($f_{\text{acc}}$), and number of jets ($N_{\text{jet}}$).
  Relative statistical ($\sigma_{\text{stat}}$) and systematic ($\sigma^{\text{tot}}_{\text{syst}}$) uncertainties in $\langle N_{\Xi} \rangle$ are listed as percentages. The total systematic uncertainty $\sigma^{\text{tot}}_{\text{syst}}$ is decomposed into contributions from the $\Lambda$-mass selection ($\Delta m_{\Lambda}$), the signed $\Xi^{-}$ decay radius in the $R\phi$ plane ($R_{\Xi}$), the method used to extract the signal yield ($\text{Fit}$), and the acceptance correction ($\text{Acc.}$). The individual uncertainties are added in quadrature to obtain the total uncertainty ($\sigma_{\text{total}}$) listed in the last column.
  }
  \label{tab:xi_yield_by_jet_rank}
  \small
  \setlength{\tabcolsep}{3.0pt}
  \begin{tabular}{@{}cc
       S[table-format=4.1, round-precision=1, scientific-notation=false]
       S[table-format=2.1, round-precision=1, scientific-notation=false]
       S[table-format=2.1, round-precision=1, scientific-notation=false]
       S[table-format=7.0, round-precision=0, scientific-notation=false]
       S[table-format=2.2, round-precision=1, scientific-notation=false]
       c
       S[table-format=2.1, round-precision=1, scientific-notation=false]
       S[table-format=2.1, round-precision=1, scientific-notation=false]
       S[table-format=2.1, round-precision=1, scientific-notation=false]
       S[table-format=2.1, round-precision=1, scientific-notation=false]
       S[table-format=2.1, round-precision=1, scientific-notation=false]
       c@{}}
    \toprule
    \multirow{2}{*}{Jet Num.} &
    \multirow{2}{*}{Jet Rank} &
    \multicolumn{1}{c}{\multirow{2}{*}{$N_{\text{obs}}$}} &
    \multicolumn{1}{c}{$\epsilon$} &
    \multicolumn{1}{c}{$f_{\text{acc}}$} &
    \multicolumn{1}{c}{\multirow{2}{*}{$N_{\text{jet}}$}} &
    \multicolumn{1}{c}{$\langle N_{\Xi}\rangle$} &
    \multicolumn{1}{c}{$\sigma_{\text{stat}}$} &
    \multicolumn{4}{c}{$\sigma_{\text{syst}}$} &
    \multicolumn{1}{c}{\multirow{2}{2.5em}{\centering $\sigma_{\text{syst}}$}} &
    \multicolumn{1}{c}{\multirow{2}{2.5em}{\centering $\sigma_{\text{total}}$}} \\
    \cmidrule(lr){9-12}
    & & &
    \multicolumn{1}{c}{$[\%]$} &
    \multicolumn{1}{c}{$[\%]$} &
    \multicolumn{1}{c}{} &
    \multicolumn{1}{c}{$\times 10^{-3}$} &
    \multicolumn{1}{c}{$[\%]$} &
    \multicolumn{1}{c}{$\Delta m_{\Lambda}$} &
    \multicolumn{1}{c}{$R_{\Xi}$} &
    \multicolumn{1}{c}{Fit} &
    \multicolumn{1}{c}{Acc.} &
    \multicolumn{1}{c}{$[\%]$} &
    \multicolumn{1}{c}{$[\%]$} \\
    \midrule
    \addlinespace
    3-jet & Jet 1 & 760.5 & 10.4 & 84.5 & 889570 & 9.71 & \multicolumn{1}{c}{$^{+6.7}_{-6.7}$} & 2.9 & 4.7 & 0.5 & 7.8 & 9.5 & $^{+11.7}_{-11.6}$ \\
    3-jet & Jet 2 & 789.4 & 10.3 & 88.0 & 889570 & 9.74 & \multicolumn{1}{c}{$^{+6.7}_{-6.7}$} & 0.5 & 1.7 & 2.1 & 6.0 & 6.6 & $^{+9.4}_{-9.4}$ \\
    3-jet & Jet 3 & 852.0 & 13.8 & 92.2 & 889570 & 7.54 & \multicolumn{1}{c}{$^{+5.8}_{-5.7}$} & 2.8 & 6.8 & 2.1 & 3.9 & 8.6 & $^{+10.3}_{-10.3}$ \\
    \bottomrule
  \end{tabular}
\end{table*}

\begin{figure}[t]
    \centering
    \includegraphics[width=0.7\linewidth]{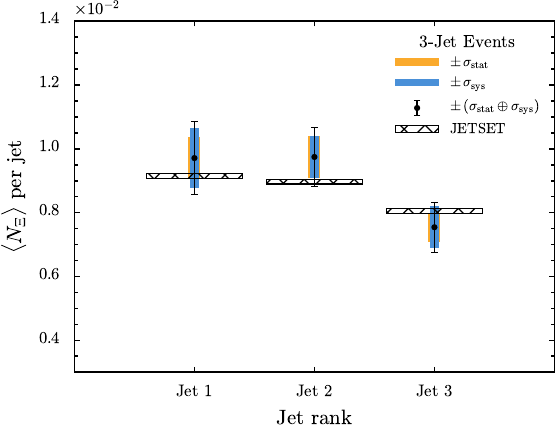}
    \caption{Average $\Xi^{-}$ production per jet in the energy-ranked jets for (left) two- and (right) three-jet events.
    The measurements are compared with the JETSET expectation.
    }
    \label{fig:unc_rank}
\end{figure}

\paragraph*{\textbf{Result}} Figure~\ref{fig:unc_rank} shows the average $\Xi^{-}$ baryon
production per jet for energy-ranked jets, and the corresponding values are given in
Table~\ref{tab:xi_yield_by_jet_rank}. 
The measurements in the leading and subleading jets are similar in both two- and three-jet events.
In three-jet events, the softest jet category gives a lower value.
All measurements remain compatible with the JETSET prediction within the uncertainties.

\begin{figure*}[t!]
    \centering
    \includegraphics[width=0.3\linewidth]{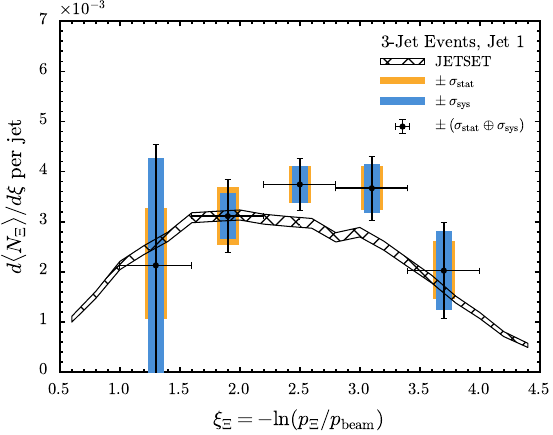}
    \includegraphics[width=0.3\linewidth]{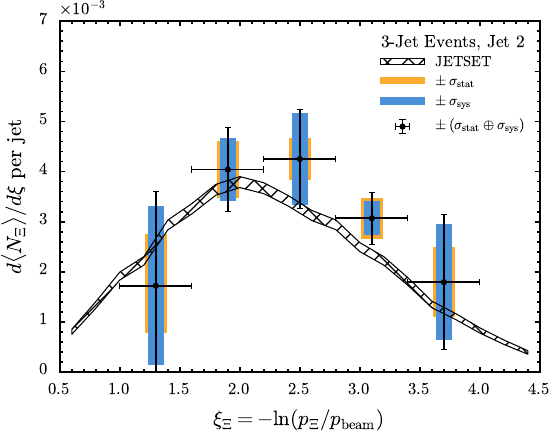}
    \includegraphics[width=0.3\linewidth]{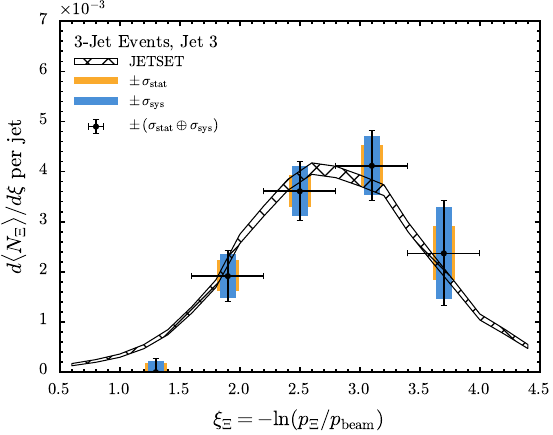}
    \caption{Differential $\Xi^{-}$ production spectra as a function of $\xi_{\Xi}=-\ln(p_{\Xi}/p_{\mathrm{beam}})$ 
    in \textbf{\textit{Jet 1}}, \textbf{\textit{Jet 2}}, and \textbf{\textit{Jet 3}} of three-jet events.}
    \label{fig:unc_rank_xi}
\end{figure*}

\begin{figure}[b!]
    \centering
    \includegraphics[width = 0.98\linewidth]{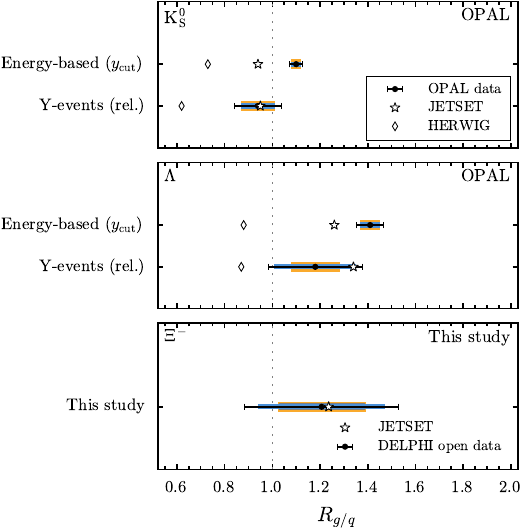}
    \caption{Comparison of the gluon-to-quark jet production ratio $\mathcal{R}_{g/q}$ for strange hadrons. The $K_S^0$ and $\Lambda$ results, along with the JETSET and HERWIG predictions, come from the OPAL measurements~\cite{OPAL:1998izp}.
    Yellow and blue bands indicate statistical and systematic uncertainties, and black error bars show their quadrature sum.}
    \label{fig:rgq}
\end{figure}

\section{Momentum spectra}
The $\xi = -\ln(p_{\Xi}/E_{\text{beam}})$ spectra of the $\Xi^{-}$ baryons in energy-ranked jets are also measured in this study.
Within each energy-ranked jet category, the selected $\Xi^{-}$ candidates are further divided into bins of $\xi$.
\begin{equation}
    \frac{\mathrm{d}\left\langle N_{\Xi}\right\rangle}
         {\mathrm{d}\xi}
    \simeq
    \frac{\Delta\left\langle N_{\Xi}\right\rangle}
         {\Delta\xi}
    =
    \frac{1}{\Delta\xi}
    \frac{1}{N_{\text{jet}}}
    \frac{N_{\text{obs}}}{\epsilon}
    \label{eq:observable_xi_spectrum}
\end{equation}

The efficiency correction and uncertainty estimation follow the procedure described in Section~\ref{sec:analysis}. The results are shown in Figure~\ref{fig:unc_rank_xi}. The average $\xi$ values for the $\Xi^{-}$ baryons in \textbf{\textit{Jet 1}}, \textbf{\textit{Jet 2}}, and \textbf{\textit{Jet 3}} are \aveXiJetA, \aveXiJetB, and \aveXiJetC, respectively. The momentum of the $\Xi^{-}$ baryons in \textbf{\textit{Jet 3}} tends to be lower than that in \textbf{\textit{Jet 1}} and \textbf{\textit{Jet 2}}. The spectra are consistent with the JETSET prediction.

\section{Relative $\Xi^{-}$ production and gluon-to-quark rate}

% The energy ordering of jets in 3-jet events provides a practical way to form quark- and gluon-enriched jet samples. 
% The interpretation, however, is not a
% direct tagging of pure quark and gluon jets.
Following the methods of previous studies~\cite{OPAL:1993uun, OPAL:1998izp}, this study determines $\mathcal{R}^{\Xi^{-}}_{g/q}$ as defined in Eq.~\ref{eq:Rgq}.

The jet samples are treated as mixtures of gluon and quark jets.
For a jet category indexed by $k$, the observed production rate of an identified particle ($h$) relative to the production of inclusive charged tracks is written as
\begin{equation}
    \mathcal{R}^{h}_{k} \equiv \frac{\langle N_{h}\rangle}{\langle N_{\text{ch}}\rangle}= \rho_{k}\cdot \mathcal{R}_{g}+ (1-\rho_{k})\cdot \mathcal{R}_{q},
\end{equation}
where $\langle N_{h}\rangle$ is the average yield of the particle per jet, $\langle N_{\text{ch}}\rangle$ is the average yield of inclusive charged tracks, and $\rho_{k}$ is the gluon fraction of the jet category.
The $\mathcal{R}_{q}$ and $\mathcal{R}_{g}$ denote the relative production rates of the particle in quark and gluon jets, respectively.
The $\mathcal{R}_{g}$ and $\mathcal{R}_{q}$ can be obtained by minimizing
\begin{equation}
    \chi^2 =
    \left(\mathbf{R}^{h}-A\boldsymbol{\theta}\right)^T
    V^{-1}
    \left(\mathbf{R}^{h}-A\boldsymbol{\theta}\right),
\end{equation}
where $V$ is the covariance of $\mathbf{R}^{h}$.
The ratio of the production rates for pure gluon and quark jets may be determined as $\mathcal{R}_{g/q} \equiv \mathcal{R}_{g}/\mathcal{R}_{q}$.

In this study, $\langle N_{\text{ch}}\rangle_{k}$ is measured in the same events and energy-ranked jet categories
as the numerator of $\mathcal{R}^{\Xi^{-}}_{k}$.
The corresponding reconstruction efficiency is evaluated from simulations as the ratio of reconstructed to generator-level charged-particle yields and is used to correct the multiplicity measured in data.

Taking $\langle N_{\text{ch}} \rangle_{k}$ and $\langle N_{\Xi} \rangle_{k}$ as inputs,
the fit gives an effective gluon-to-quark ratio for $\Xi^{-}$ production of $\mathcal{R}^{\Xi^{-}}_{g/q} = \RgqCentral \pm \RgqStat~\mathrm{(stat.)} \pm \RgqSys~\mathrm{(syst.)}$.
The uncertainties in $\langle N_{\mathrm{ch}}\rangle_k$ and $\langle N_{\Xi}\rangle_k$ are propagated to the fitted ratio, while correlated systematic uncertainties in $\langle N_{\mathrm{ch}}\rangle_k$ largely cancel.

The nominal gluon fractions are obtained from the official MC simulation.
To estimate the effect of the MC generator on the $\mathcal{R}^{\Xi^{-}}_{g/q}$ result, a Pythia8 sample with a sample size comparable to that of the data is used, providing alternative gluon fractions of \gluonFracJetAAlt, \gluonFracJetBAlt, and \gluonFracJetCAlt for the three energy-ranked jet categories, respectively.
Repeating the fit with these fractions changes $\mathcal{R}^{\Xi^{-}}_{g/q}$ by an amount much smaller than the other uncertainties.

Figure~\ref{fig:rgq} compares this result with the $\mathcal{R}_{g/q}$ values for $K_{S}^{0}$ and $\Lambda$ production measured by OPAL~\cite{OPAL:1998izp}.

\section{Conclusion}

This work presents the first measurement of the production of doubly strange baryons, $\Xi^{-}/\bar\Xi^{+}$, in gluon- and quark-enriched jets using three-jet events from hadronic $Z$ boson decays recorded by the DELPHI experiment during 1992--1995.
Jets are reconstructed with the Durham 
algorithm using $y_{\text{cut}} = 0.005$ and are ordered according to their energies. The lowest-energy jet is enriched in gluon jets, while the two higher-energy jets are enriched in quark jets.
The softest jets are observed to have lower $\Xi^{-}/\bar{\Xi}^{+}$ production rates and softer momentum spectra than the other two energy-ranked jets.
These results are consistent with the JETSET prediction.

The ratio of $\Xi^{-}/\bar{\Xi}^+$ production rates in gluon and quark jets, each normalized to the corresponding mean charged-particle multiplicity, is determined to be $\mathcal{R}^{\Xi^{-}}_{g/q} = \RgqCentral \pm \RgqStat~\mathrm{(stat.)} \pm \RgqSys~\mathrm{(syst.)}$, consistent with $\mathcal{R}^{K_S^0}_{g/q}$ and $\mathcal{R}^{\Lambda}_{g/q}$ reported by OPAL~\cite{OPAL:1998izp}.
The $\mathcal{R}^{\Xi^{-}}_{g/q}$ ratio is consistent with unity within uncertainties, indicating that the enhancement of $\Xi^{-}/\bar{\Xi}^{+}$ production in gluon jets relative to quark jets is compatible with that observed for inclusive charged particles.
The precision of the present measurement is limited by the available data sample. Future $e^{+}e^{-}$ colliders such as CEPC and FCCee operating at the $Z$ pole are expected to provide substantially larger data samples which will enable more precise studies of productions of hadrons in gluon and quark jets.

\begin{acknowledgments}
The authors are grateful to the DELPHI Collaboration for the open data which allows this study to be carried out. We thank the BESIII Collaboration and Xiongfei Wang for sharing the analysis code and for their assistance with the vertex-fitting technique. We also thank Hideki Okawa for valuable suggestions.

\end{acknowledgments}

% \clearpage

\nocite{*}

\bibliography{references}% Produces the bibliography via BibTeX.

\end{document}